\documentclass[aps,preprint,showpacs,superscriptaddress,groupedaddress]{revtex4-2}  
\usepackage{graphicx}  
\usepackage{dcolumn}   
\usepackage{bm}        
\usepackage{amssymb}   
\usepackage{amsmath}
\usepackage[colorlinks=true, citecolor=blue, urlcolor=blue, linkcolor = blue]{hyperref}
\usepackage{bm}
\usepackage{color}
\usepackage{bookmark}
\usepackage{tabularx}
\usepackage{mathtools}
\usepackage{microtype}
\usepackage{cancel}
\usepackage{relsize}
\usepackage{float}
\usepackage[caption = false]{subfig}
\usepackage{xcolor}

\begin{document}

%

\title{Dicke model semiclassical dynamics in superradiant dipolar phase in the 'bound luminosity' state}


\author{S. I. Mukhin}
\author{A. Mukherjee}
\author{S.S. Seidov}
\affiliation{Theoretical Physics and Quantum Technologies Department, NUST ``MISIS'', Moscow, Russia}


\begin{abstract}
Analytic solution of semiclassical dynamics equations of the Dicke model in a superradiant state is presented. The time dependences of the amplitudes of superradiant bosonic condensate and coherent two-level atomic array in the microwave cavity prove to be expressed via Jacobi elliptic functions of real time and manifest existence of an adiabatic invariant of motion in the strongly coupled system. The periodic beatings of the photonic and atomic coherent state amplitudes are shifted in time revealing an effect of 'bound luminosity', when energy stored in the two-level system during 'darkness' in the cavity is suddenly converted into photonic condensate that 'illuminates' the cavity for half a period, before it plunges into 'darkness' again.    
\end{abstract}

\maketitle

 \section{Introduction }
 Predicted previously superradiant quantum phase transition in the Dicke model \cite{Dicke} of an array of $N\gg1$ two-level systems (TLS) coupled to a single bosonic mode in the resonant cavity \cite{Brandes, Brandes_entlg} was shown recently \cite{MukhinPRA} to map on the self-consistent rotation by a finite angle of the Holstein-Primakoff \cite{HP, Karassiov} representation (RHP) of the pseudo-spin operators describing coherent states of the TLS.  
We consider the Dicke model \cite{Dicke} Hamiltonian in the following form: 

\begin{align}
\hat{H} =  \frac{1}{2}\left(\hat{p}^2+\omega^2 \hat{q}^2 \right) + g \hat{p} \, \hat{S}^y  - E_J\, \hat{S}^z \label{UHU_CPB},
\end{align}
where we have introduced operators $\hat{S}^\alpha=\sum_i \hat{s}_i^\alpha$ of  the total pseudo-spin components describing the TLS. For a definiteness one may consider an array of $N$ superconducting islands, each divided into two sub-islands by a Josephson junction (JJ) taken in a Cooper pair box (CPB) condition, see e.g. \cite{Shnirman}. The array is placed inside the resonant cavity with $\omega$ being the frequency of a single photonic mode.
The photonic field second quantized operators are: 
\begin{align}
\hat{p}=\mathrm{i} \sqrt{\frac{\hbar\omega}{2}}\left(\hat{a}^\dag -\hat{a} \right)\;\;\; \text{and} \;\;\;  \hat{q}= \sqrt{\frac{\hbar}{2\omega}}\left(\hat{a}^\dag + \hat{a} \right)\, \label{pq},
\end{align}
where $\left[\hat{a},\hat{a}^\dag\right]=1$.
The total spin $\hat{S}^2$ is conserved, because it commutes with (\ref{UHU_CPB}): $\left[\hat{S}^2, H\right]=0$.
Cooper pairs tunnelling is represented by $-E_J \hat{S}^z$ term. The term $g\hat{p}\,\hat{S^y}$ describes a dipole coupling  between Cooper pair box and photonic field with the coupling strength $g$.  In \cite{Brandes} it was demonstrated that a quantum phase transition sets the system in (\ref{UHU_CPB}) into a double degenerate dipolar ordered state.
 Here we provide the analytical solutions describing dynamics of JJ in a resonant cavity demonstrating metastable 'bound luminosity' state with periodic coherent beatings  of the dipoles, that connect  the two double degenerate dipolar ordered phases discovered previously.
The first two terms in (\ref{UHU_CPB}) express the energy of the harmonic oscillator of a single photonic mode. Energy splitting between the states of the TLS subsystem is described by  $-E_J \hat{S}^z$ term, whereas  $g\hat{p}\hat{S^y}$ represents a dipole coupling energy of the TLS with photons. It was demonstrated in \cite{MukhinPRA}, that introduction of self-consistently rotating Holstein-Primakoff representation (RHP)  for the operators $\hat{S}_{x,y,z}$ provides a rather transparent description of the second order quantum phase transition in Dicke model (\ref{UHU_CPB}). Namely, when the coupling constant $g$ becomes greater than a critical value $g_c$ the superspin $S=N/2$ gradually rotates from $z$-axis to $y$-axis by an angle $|\theta|\leq \pi/2$, where the upper limit corresponds to the maximum of the total 'dipole moment' $\langle\hat{S}_{y} \rangle \propto S\sin{\theta}$ of the TLS. Simultaneously, photonic field operator $ \hat{p}$ acquires an off-diagonal mean value $\langle\hat{p} \rangle \propto -gS\sin{\theta}$, manifesting formation of the macroscopically coherent ('superradiant') photon condensate coupled to dipole moment of the TLS with an energy $\propto g\langle\hat{S}_{y} \rangle \langle\hat{p} \rangle $. Obviously, simultaneous change of signs:  $\langle\hat{S}_{y} \rangle ,\langle\hat{p} \rangle \rightarrow$ $-\langle\hat{S}_{y} \rangle ,-\langle\hat{p} \rangle $ brings the system into the new state with the same spin-photon coupling energy, thus, manifesting degeneracy of the ground state.
Here we present analytical solutions of the equations that describe "slow" meandering of the system between the degenerate ground states: the energy stored in TLS $\sim \langle S_z \rangle $ is periodically pumped into photonic condensate, while the amplitude of the total dipole moment of the TLS $\sim \langle S_y \rangle $  and the amplitude of the superradiant condensate $\sim \langle{p} \rangle $ change their signs for opposite ons as functions of time. We found two integrals of motion for these slow motion of coherent amplitudes in the semiclassical limit.  


The equations of the semi-classical dynamics of the ground state expectation values of the photon momentum and  Cartesian components of the total pseudo-spin could be derived along the line described in \cite{MukGnezd} starting from the Heisenberg equation for the quantum operators written as:

\begin{align}
 \ddot{\hat{A}}=-\frac{1}{\hbar^2}\left[\hat{H},\left[\hat{H},\hat{A}\right]\right]\,.
\label{Heisenberg}
\end{align}
\noindent
where $\hat{A}=\hat{p},\hat{S}_{\alpha}$ and $\alpha=x,y,z$. The next step is to use Dicke Hamiltonian (\ref{UHU_CPB}) in (\ref{Heisenberg}), allowing for commutation relations for the spin components and harmonic oscillator coordinate and momentum. Then, one obtains the following system of differential equations: 
 \begin{align}
&{}\ddot{\hat{S}}_z=-{g}^2\hat{p}^2\hat{S}_z-gE_J\hat{p}\hat{S}_y\,\label{sz_dyn}\\
&{}\ddot{\hat{S}}_y=-{E_J}^2\hat{S}_y-gE_J\hat{p}\hat{S}_z\,\label{sy_dyn}\\
&{}\ddot{\hat{S}}_x=-{E_J}^2\hat{S}_x-g^2\hat{p}^2\hat{S}_x\,\label{sx_dyn}\\
&{}\ddot{\hat{p}}=-\omega^2\left\{\hat{p}+g\hat{S}_y\right\}\, \label{p_dyn}\,
\end{align}
In order to pass from operators to their ground state semiclassical expectation values in the superradiant state the following steps are made. First, the amplitude of superradiant photonic condensate $\lambda_R$ is introduced \cite{Brandes} as a shift of the Bose-operators in (\ref{pq}):

\begin{align}
&\hat{a}^\dag = \hat{c}^\dag - \mathrm{i}\lambda_R;\; \hat{a} = \hat{c} + \mathrm{i}\lambda_R,\\
&\hat{p}= \sqrt{2\hbar\omega}\lambda_R+\mathrm{i} \sqrt{\frac{\hbar\omega}{2}}\left(\hat{c}^\dag -\hat{c} \right);\;\hat{q}= \sqrt{\frac{\hbar}{2\omega}}\left(\hat{c}^\dag + \hat{c} \right)\, \label{a_shift}.
\end{align}

As was found in \cite{MukhinPRA}, simultaneously with developing photonic condensate $\lambda_R$ the system rotates its total pseudo-spin by an angle $\theta$ around x axis. Here we rotate the total spin by an angle $\theta$ around x axis and then by an angle $\phi - \pi/2$ around z axis, where $\theta$,  $\phi$ are time dependent:

\begin{align}
\begin{cases} \hat{S}^{z}=\hat{J}^z \cos \theta - \hat{J}^y \sin \theta 
\\ \hat{S}^{y}=\hat{J}^z \sin \theta \sin \phi + \hat{J}^y \cos \theta \sin \phi - \hat{J}^x \cos \phi 
\\ \hat{S}^{x}=\hat{J}^z \sin \theta \cos \phi + \hat{J}^y \cos \theta \cos \phi + \hat{J}^x \sin \phi \label{S}\end{cases}
\end{align} 

\noindent where a set of operators of the Cartesian projections of the total spin $\hat{J}^{x,y,z}$ is defined using Holstein-Primakoff relations: 

\begin{align}
\begin{cases}\hat{J}^z=S-\hat{b}^\dag \hat{b} \\  
\hat{J}^y=\mathrm{i}\displaystyle\sqrt{\frac{S}{2}} \left( \hat{b}^\dag \sqrt{1-\frac{\hat{b}^\dag \hat{b}}{2S}} - \sqrt{1-\frac{\hat{b}^\dag \hat{b}}{2S}} \, \hat{b} \right)
\simeq \mathrm{i}\displaystyle\sqrt{\frac{S}{2}} \left( \hat{b}^\dag  - \hat{b} \right) \\ 
\hat{J}^x=\displaystyle\sqrt{\frac{S}{2}} \left( \hat{b}^\dag \sqrt{1-\frac{\hat{b}^\dag \hat{b}}{2S}} + \sqrt{1-\frac{\hat{b}^\dag \hat{b}}{2S}} \ \hat{b}
\right) \simeq \sqrt{\frac{S}{2}} \left( \hat{b}^\dag  + \hat{b} \right)\end{cases} \ \label{HP_Dicke},
\end{align}
and $\left[\hat{b},\hat{b}^\dag\right]=1$.

The reason for a rotation is simple: in the ground state with a small coupling strength $g$ the angle is: $\theta=0$, and operator $\hat{S}_z=\hat{J}^z$ in DH (\ref{UHU_CPB}) is approximately diagonal and, hence, diagonal element of $\langle \hat{S}^{y} \rangle = \langle\hat{J}^y \rangle $ is vanishing. Thus, "dipole moment" $\propto \langle\hat{S}^{y} \rangle $ is zero and photons are semiclassically decoupled form the TLS subsystem. On the other hand, when rotation angle $\theta$ becomes not vanishing, a finite "dipole moment" of the TLS subsystem $\propto \langle\hat{S}^{y} \rangle =\langle\hat{J}^z  \rangle \sin \theta$ emerges and couples to the coherent photon condensate $\langle\hat p \rangle =\lambda_R$. The latter  makes the bare ground state of the TLS subsystem unstable. To proceed, we substitute operators in equations (\ref{sz_dyn})-(\ref{p_dyn}) with their ground state expectation values in the Heisenberg representation: $\hat{A}\rightarrow \langle\hat{A} \rangle $ using relations (\ref{a_shift}), (\ref{S}) and (\ref{HP_Dicke}). In what follows, we consider the resonant case: $\hbar\omega=E_J$. 
 
Contrary to \cite{Brandes}, no shift of the Holstein-Primakoff bosons in (\ref{HP_Dicke}) is made, since its role is taken by rotation angle $\theta$. It is obvious, then, from (\ref{sz_dyn})-(\ref{sx_dyn}) that the system of dynamic equations describes nonlinear evolution of the total spin $\vec{S}$ in a frame rotating around the axis $z$  in the  pseudo-spin space with the frequency $E_J$ . Since we consider the resonant case $\omega=E_J$, the angle $\phi$ appearing in (\ref{S}) changes with time as $\phi = \omega t$, which means the total spin $\vec{S}$ rotates  with a constant angular velocity $\omega$ around $z$ axis, and photonic variable $p$ in (\ref{p_dyn}) oscillates with the frequency $\omega$ as well, but is linearly coupled to the spin via the last term in the rhs of (\ref{p_dyn}). Then, it is convenient to introduce the following representation for the expectation values as functions of time ($\hbar=1$):
 \begin{align}
&{}p= \sqrt{2\omega}\lambda_R(t)\equiv -\sqrt{2\omega}g_0(t)\cos{\omega t}\,\label{lam_t}\\ 
&{}{S_z}=S\cos\theta(t) \label{z_Rt}\\
&{}{S_y}=S\sin\theta(t)\sin{\omega t}\label{y_It}\\
&{}{S_x}=S\sin\theta(t)\cos{\omega t}\label{x_It}\,,
\end{align}
\noindent
Since in the superradiant state\cite{Brandes,MukhinPRA} photonic shift is $\lambda_R\propto S=N/2 \gg 1$ in thermodynamic limit , one can use semiclassical approximation for the ground state : $\langle\hat{p}^2 \rangle  \approx \langle\hat{p} \rangle ^2$ in (\ref{sz_dyn}) - (\ref{p_dyn}). 

\section{Dipolar phase transition with 'frozen' crossed electric and magnetic fields}

It is remarkable, that in the static limit, i.e. assuming  ${\lambda}_R,\,\theta,\,\phi \equiv\text{const}$ in the definitions in the above (\ref{lam_t}) - (\ref{x_It}), 
three equations (\ref{sz_dyn}),(\ref{sy_dyn}) and (\ref{p_dyn}) become algebraic and collapse explicitly into two equations,
since equation (\ref{sz_dyn}) becomes an identity. This result indicates that double degenerate solutions signify phase transition into spontaneously ordered dipolar state of the TLS array with the {\it{frozen crossed electric and magnetic fields}} inside the resonant cavity. 
It is apparent, that such ground state of the system might be possible if  the persistent crossed  electric and magnetic fields obey static conditions on the cavity's boundary, i.e. both surface static charge
polarisation and constant (super)current density should be allowed. We also emphasise here, that as it is apparent from the right-hand sides of the equations (\ref{sy_dyn}) and (\ref{p_dyn}), there are indeed two
static solutions that differ by the simultaneous sign change: $\theta\rightarrow -\theta$ and ${\lambda}_R\rightarrow - {\lambda}_R$ (can be compared with \cite{Brandes}).

\section{``Bound luminosity'' of superradiant state in the Dicke model in the adiabatic TLS limit}

Now we derive analytical solutions of the dynamics equations (\ref{sz_dyn})-(\ref{p_dyn}) using a two time scale approach, see also \cite{MukGnezd}, in the adiabatic limit of the TLS evolution. We neglect Eq. (\ref{sx_dyn}) because  $\hat{S_x}$ does not enter Dicke Hamiltonian (\ref{UHU_CPB}) as well as equations (\ref{sz_dyn}),(\ref{sy_dyn}) and (\ref{p_dyn}). Also we split the time-dependence into 'fast' rotation $\propto \exp(i\omega t)$, and 'slow' evolution described by the functions $g_0(t)$ and $\theta(t)$ defined by relations (\ref{lam_t})-(\ref{x_It}). Then, substituting (\ref{lam_t})-(\ref{x_It}) into (\ref{sy_dyn}) and (\ref{p_dyn}), and averaging left and right hand sides of the equations over fast oscillations with the frequency $\omega$,
one obtains the following system of equations for the adiabatically changing functions: 
\begin{align}
&{}\dot{\theta}(t)=\frac{g\sqrt{2\omega}}{2}\,g_0(t)\, \label{theta_t}\\ 
&{}\sin\theta=-\frac{2\sqrt{2}}{gS\sqrt{\omega}}\dot{g}_0(t)\,\label{g0_T}
\end{align}
\noindent and equation (\ref{sz_dyn})  becomes an identity in the adiabatic (quasi static) limit and drops out as it did in the static case described above. 
The procedure of averaging over fast oscillations, that  leads to (\ref{theta_t}) and (\ref{g0_T}) is simple. Namely, one substitutes (\ref{lam_t})-(\ref{x_It}) in (\ref{sy_dyn}) and then multiplies both sides of the equation by $\cos{(\omega t)}$. Then, take average over 'fast' oscillations over a short time period ($\sim \omega^{-1}$), while considering variables containing 'slow' time scale (adiabatic TLS evolution)  as constants. After integration over the short time period one finds (\ref{theta_t}). Analogously, multiplying both sides of equation (\ref{p_dyn}) by $\sin{(\omega t)}$ and repeating the same procedure one finds (\ref{g0_T}). By the same token, Eq. (\ref{sz_dyn}) turns out into identity like in the 'frozen' case considered above, after one neglects the second time derivative of the 'slow' function $\cos\theta$ in the left hand side of (\ref{sz_dyn}).
In what follows, the natural unit convention is used, $\hbar=1$, for the sake of simplicity of notations.  Multiplying left hand side of (\ref{theta_t}) by $\sin\theta$ and using (\ref{g0_T}) one obtains a complete differential that gives the first adiabatic integral of the fast oscillations averaged dynamic equations:
\begin{align}
\cos\theta(t)-\frac{{{g}^2}_0(t)}{S}=\text{const}\equiv C\, \label{integral_1},
\end{align}
where $C$, within a constant factor, is the total fast oscillations averaged energy of the system: $C\propto \langle \hat{H}\rangle$. Next, differentiating (\ref{g0_T}) once with respect to time  and expressing $\cos\theta(t)$ via $g_0(t)$ using (\ref{integral_1}), one finally obtains a nonlinear second order differential equation for unknown function $g_0(t)$ in the closed form:
\begin{align}
\frac{2\ddot{g}_0(t)}{gS\sqrt{\omega}}+\frac{g\sqrt{\omega}}{2}g_0(t)\left(C+\frac{{{g}^2}_0(t)}{S}\right)=0. \    \label{cn}
\end{align}
This equation is exactly solvable in terms of Jacobi elliptic functions. If we choose $C = 1 - 2 k^2$ with $0\leq k\leq 1$ \cite{Witt}, then solution of (\ref{cn}) is given by:
\begin{align}
{g}_0(t)=\sqrt{2S}k\, cn\left(\frac{g\sqrt{S\omega}}{2}\,t\,,k\right) , \ \  \lambda_R =  \sqrt{2S}k\, cn\left(\frac{g\sqrt{S\omega}}{2}\,t\,,k\right) \cos{\omega t}  \label{cn_sol1}
\end{align}
where $k$ is called the elliptic modulus. 
Using solution (\ref{cn_sol1}) the other variables can be found algebraically from relations (\ref{theta_t}), (\ref{g0_T}) as (compare \cite{MukGnezd}): 
\begin{align}
 &{} S_z = S\cos{\theta} = S\{1 -2 k^2 \ sn^2\left(\frac{g\sqrt{S\omega}}{2}\,t\,,k\right)\}, \nonumber \\
 &{} S_y = -S\sin{\theta}\sin{(\omega t)} = 2kS \ sn\left(\frac{g\sqrt{S\omega}}{2}\,t\,,k\right)dn\left(\frac{g\sqrt{S\omega}}{2}\,t\,,k\right)\sin{\omega t}.\label{zRzI1}
\end{align}
On the other hand, if we choose $C = k^2 - 2,$ then the set of solutions would be as follows: 
\begin{align}
{g}_0(t)=\sqrt{2S}\, dn\left(\frac{g\sqrt{S\omega}}{2}\,t\,,k\right) , \ \  \lambda_R =  \sqrt{2S} \, dn\left(\frac{g\sqrt{S\omega}}{2}\,t\,,k\right) \cos{\omega t},  \label{cn_sol2}
\end{align}
\begin{align}
 &{} S_z = S\cos{\theta} = Sk^2 \{ 1 -2  \ sn^2\left(\frac{g\sqrt{S\omega}}{2}\,t\,,k\right)\}, \nonumber \\
 &{} S_y = -S\sin{\theta}\sin{(\omega t)} =  2k^2 \ sn\left(\frac{g\sqrt{S\omega}}{2}\,t\,,k\right)cn\left(\frac{g\sqrt{S\omega}}{2}\,t\,,k\right) \sin{\omega t}. \label{zRzI2}
\end{align}
The effective ``frequency'' $\Omega$ characterizing beatings of the Jacobi elliptic functions in (\ref{cn_sol1})-(\ref{zRzI2}) is expressed via the complete elliptic integral of the first kind $K(k)$ \cite{Witt} (diverging at $k=1$) in the following way:

\begin{align}
{g}_0(t+2\pi/\Omega)={g}_0(t)\,,\;\; \Omega\equiv \omega\frac{\pi}{K(k)}\frac{g}{2g_\mathrm{c}}\,, \;\; g_\mathrm{c}\equiv \sqrt{E_J/S}.  \label{Omega_k}
\end{align}
Here $g_\mathrm{c}$ is critical coupling strength above which the superradiat state is stable \cite{Brandes}. Hence, in the limit $k\rightarrow 1$,  the 'slow' evolution assumption of $\theta(t)$, that is $\Omega\ll \omega$, which we have assumed at the beginning of the calculation
is satisfied if:
\begin{align}
\Omega\equiv \omega\frac{\pi}{K(k)}\frac{g}{2g_\mathrm{c}}\ll\omega\;,\;\;\text{i.e.}\;\; K(k)\gg\frac{\pi}{2}\frac{g}{g_\mathrm{c}}\;,\;\text{hence}\,,\;\;1-2\displaystyle\exp\left\{-\frac{\pi}{2}\frac{g}{g_\mathrm{c}}\right\} \leq k\leq 1.\,\label{assump_k}
\end{align}
 Thus, a remarkable physical picture of the ``bound luminosity'' phase is described by the solutions (\ref{cn_sol1})-(\ref{zRzI2}). 
 The expressions of electric field and total dipole moment generated inside the resonant cavity, that are meandering between the two degenerate ground states of the Dcke system, are given as:

\begin{align}
\hat{\vec{E}}= \mathrm{i} \sqrt{\frac{1}{V}} \left(\hat{a}^\dag-\hat{a}\right) \vec{\epsilon}\, \label{E}
\end{align}
\begin{align}
\hat{\vec{d}} = - 2 e \, l\vec{\epsilon} \hat{S}_y   \label{dipole_total},
\end{align}
 where $\vec{\epsilon}, V, l$ are the polarization vector, volume of the cavity and effective thickness across a JJ respectively. $2e$ is the elementary charge of the Cooper's pair. 
 Now, for the two choices of the adiabatic invariant  $C$, which we have discussed above, the following two sets of solutions for the electric field $\vec{E}$ and the total TLS dipole moment $\vec{d}$ are:
 
 \begin{enumerate}
 
  \item {\bf Case 1 : $C = 1 - 2 k^2$}
 
\begin{align}
&{}\vec{E}(t)=\frac{2\sqrt{\omega}}{\sqrt{V}}\,\vec{\varepsilon}\,\sqrt{2S}k\,cn\left(\frac{g\sqrt{S\omega}}{2}\,t\,,k\right)\cos(\omega\,t)  \label{E_t1}\\ 
&{}\vec{d}(t)=2elS\vec{\varepsilon}2k\,sn\left(\frac{g\sqrt{S\omega}}{2}\,t\,,k\right)\,dn\left(\frac{g\sqrt{S\omega}}{2}\,t\,,k\right)\sin(\omega\,t)\, \label{d_t1}\\
&{}E_JS_z(t)=E_JS  \ [ 1-2k^2\,sn^2\left(\frac{g\sqrt{S\omega}}{2}\,t\,,k\right) ].\,\label{Zeeman_t1}
\end{align}

\item {\bf Case 2 : $C =  k^2 - 2 $}

\begin{align}
&{}\vec{E}(t)=\frac{2\sqrt{\omega}}{\sqrt{V}}\,\vec{\varepsilon}\,\sqrt{2S}\,dn\left(\frac{g\sqrt{S\omega}}{2}\,t\,,k\right)\cos(\omega\,t)  \label{E_t2}\\ 
&{}\vec{d}(t)=2elS\vec{\varepsilon}2k^2\,sn\left(\frac{g\sqrt{S\omega}}{2}\,t\,,k\right)\,cn\left(\frac{g\sqrt{S\omega}}{2}\,t\,,k\right)\sin(\omega\,t)\, \label{d_t2}\\
&{}E_JS_z(t)=E_JSk^2 \ [1-2\,sn^2\left(\frac{g\sqrt{S\omega}}{2}\,t\,,k\right)] \,\label{Zeeman_t2}
\end{align}
\end{enumerate}
\noindent
where $sn\,,dn\,,cn$ are Jacobi elliptic functions. One interesting point should be noted in this context. Namely, in both cases the expressions for TLS dipole energy in the electric field  i.e, $-\vec{E} . \vec{d}$ remain the same. There is a difference in the form of the photonic part $: \hat{p}^2/2$ and Zeeman energy part : $- E_J S_z$. But we have considered the ``slow evolution'' assumption in the derivation, which means $k \longrightarrow 1$. In this limit, the expressions of all the energy contributions for both the cases remain identical.

\begin{figure}[h!!]
\centerline{\includegraphics[width=0.5\linewidth]{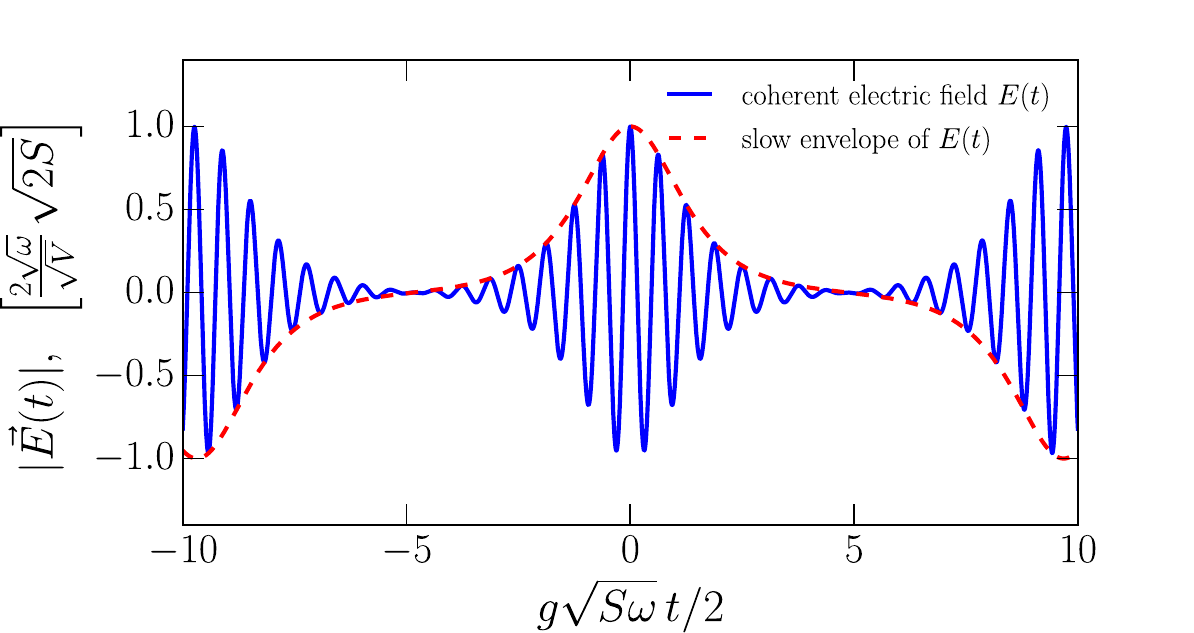}}
\caption{Time-dependent oscillations of the coherent electric field of the emerged photonic condensate (solid line) with the extracted ``slow envelope'' curve (dashed line) that manifests periodic change of the phase of the fast oscillations by $\pi$.}
\label{E_t}
\end{figure}

\begin{figure}[h!!]
\centerline{\includegraphics[width=0.5\linewidth]{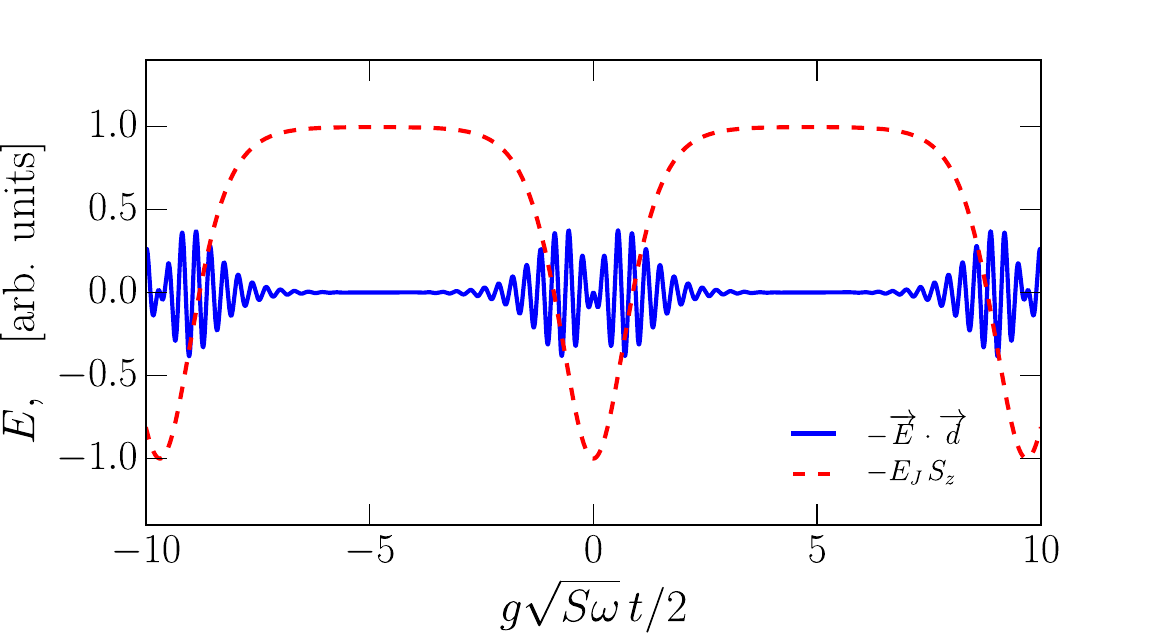}}
\caption{Time dependence of the CPB dipole energy (arb. units) in the emerged photonic condensate field $\propto -\vec{E}\vec{d}$ (solid line), and ``Zeeman'' energy (arb. units) $\propto -E_J S_z$ (dashed line). }
\label{D_Z_t}
\end{figure}

 Using above relations for the two cases, we plot  the time-dependent electric field of photonic condensate $E(t)$, as well as distribution of energy of the system in the following figures respectively. The energy is distributed between the coupling energy of the TLS dipole to the cavity electromagnetic field, $-\vec{E}\vec{d}$, and ``Zeeman'' energy, $-E_J \hat{S_z}$, i.e. the Josephson tunnelling energy of Cooper pairs that can be seen in Hamiltonian (\ref{UHU_CPB}). For $C = 1 - 2k^2$, the plot in Fig.\ref{E_t} contains time-dependent oscillations of the coherent electric field of the emerged photonic condensate (solid line) with the extracted 
``slow envelope'' curve (dashed line) that manifests periodic change of the phase of the fast oscillations by $\pi$, i.e. the double-degeneracy of the mean-field solution is indeed dynamically lifted.   The next plot in Fig.\ref{D_Z_t} clearly demonstrates that the ground state energy reveals a periodic transformation of Cooper pairs ``zero point oscillations'' energy (dashed line) into suddenly illuminated energy of photonic condensate bound to TLS dipoles  (solid line) since we have not  considered energy dissipation.

 \section{Conclusive remarks }
 In this article, we have considered the Dicke Hamiltonian of a system of $N$  half spins interacting with a single mode photonic field in a resonant cavity in the semiclassical superradiant limit.  Using a recently discovered analytical tool \cite{MukhinPRA} called rotating Holstein-Primakoff representation for the Cartesian components of the total
spin, we found the semiclassical dynamics of the system from the Heisenberg dynamics equations. Using two time ('short' and 'long') scale approach, we have derived for the first time the analytical expressions of the condensed electric field and dipole moment inside the resonant cavity. The solutions  are expressed via Jacobi elliptic functions of real time  describing  a superradiant intrinsic ``bound luminosity'' state. Many other problems related to this system like bifurcation and chaos at asymptotic limit, region of integrability etc still remain unexplored which we aim to try in future.

\section {ACKNOWLEDGMENTS}
S.I.M. acknowledges illuminating discussions with Prof. Carlo Beenakker, Prof. Konstantin Efetov, Dr. Bernard van Heck and Dr. Nikolay Gnezdilov during the course of this work, as well as hospitality of the colleagues at the Lorentz Institute during his stay in Leiden. The authors gratefully acknowledge the financial support of the Ministry of Science and Higher Education of the Russian Federation in the framework of Increase Competitiveness Program of NUST MISIS ( K4-2018-061),
implemented by a governmental decree dated 16-th of March 2013, No. 211.
%
%

\end{document}